\documentclass[doublecol]{epl2}
\title{Universality of temperature distribution in granular gas mixtures with a steep particle size distribution}
\shorttitle{Universality of granular temperature distribution}

\author{Anna Bodrova\inst{1} \and Denis Levchenko\inst{1} \and Nikolay Brilliantov \inst{2}}
\shortauthor{A. Bodrova \etal}

\institute{
  \inst{1} Faculty of Physics, M.V.Lomonosov Moscow State University, Moscow, 119991, Russia\\
  \inst{2} Department of Mathematics, University of Leicester, Leicester LE1 7RH, United Kingdom
}

\abstract{Distribution of granular temperatures in granular gas mixtures is investigated analytically and
numerically. We analyze space uniform systems in a homogeneous cooling state (HCS) and under a uniform
heating with a mass-dependent heating rate $\Gamma_k\sim m_k^{\gamma}$. We demonstrate that for steep size distributions of particles the granular temperatures obey a \emph{universal} power-law distribution, $T_k \sim
m_k^{\alpha}$, where the exponent $\alpha$ does not depend on a particular form of the size distribution, the
number of species and inelasticity of the grains. Moreover, $\alpha$ is a universal constant for a HCS and
depends piecewise linearly on $\gamma$ for heated gases. The predictions of our scaling theory agree well
with the numerical results.}

\begin{document}

\maketitle
\section{Introduction}

Granular materials, such as gravel, sand or  different types of  powders, are ubiquitous in nature and widely used in industry \cite{DryGranMed}.
Rarified granular systems, where the volume of a solid phase is small as compared to the total volume, are termed as granular gases  \cite{book}. In the Earth conditions these may be obtained by placing granular matter into a container with vibrating \cite{wildman} or rotating \cite{rotdriv} walls, applying electrostatic \cite{electrodriven} or magnetic \cite{magndriven} forces etc. Extraterrestrial granular gases are also common. Many astrophysical objects, like protoplanetary discs, Planetary Rings and interstellar dust clouds contain granular gases as one of the important component \cite{PlanRings}.

Granular particles collide inelastically, loosing their kinetic energy during  the collisions, which
transforms into heat (the energy of the internal degrees of freedom). If no  energy is supplied to a system
from an external source, the particles gradually slow down. Such state of granular gases is called a
homogeneous cooling state (HCS) \cite{book}; it can be realized in a microgravity environment,
e.g.~\cite{expgg}. In the HCS the average kinetic energy of particles decreases and after some time a space
homogeneous distribution looses its stability, e.g.~\cite{book}. Contrary, driven gases may permanently
remain homogeneous. In the present study only space uniform systems are addressed.

Real granular materials are not comprised of identical particles but rather represent a polydisperse mixture
of grains of different mass and size. Particular forms of particle size distributions may be very  different,
depending on the nature of granular systems. Power-law size distribution is, however,  rather common.  It is
observed in Planetary Rings~\cite{PlanRings} and may be used to describe industrial materials,
e.g.~\cite{Normand1986}. 

A statistical description of a space homogeneous gas is based on the velocity  distribution functions
$f_k({\bf v}_k,t)$, which gives the number of particles of sort $k$ with velocity ${\bf v}_k$ in a unit
volume at time $t$. Correspondingly,
\begin{equation}
\label{eq:nk} n_k=\int f_k\left({\bf v}_k,t\right) d{\bf v}_k
\end{equation}
are number densities of particles of the sort $k$, which keep constant in a homogeneous gas. An important
characteristic of a granular system is a mean kinetic energy of the gas particles of sort $k$, e.g.
\cite{GarzoDuftyMixture,zippelius},
\begin{equation}
\label{eq:Tk} \frac32 n_kT_k(t)=\int d{\bf v}_kf_k\left({\bf v}_k,t\right)\frac{m_kv_k^2}{2} \, ,
\end{equation}
where $m_k$ is the corresponding mass of the grains. $T_k$ are also termed "partial granular temperatures".
In a sharp contrast with molecular gases, the energy equipartition does not hold in a mixture of granular
gases, that is $T_k\neq T_l$ for $k\neq l$. This has been shown in experiments \cite{wildman} and
explained theoretically \cite{book,breyduftyHrenya,GarzoDuftyMixture,BarratTrizac2002}. Physically, this
follows from the nature of a dissipative collision between particles of different mass: The larger the mass,
the smaller the part of a particle energy lost in a collision. Quantitatively, the dissipation of energy  in
such collisions is described by a restitution coefficient:
\begin{equation}
\label{rc} \varepsilon_{ik} = \left|\frac{\left({\bf v}^{\,\prime}_{ik} \cdot {\bf e}\right)}{\left({\bf
v}_{ik} \cdot {\bf e}\right)}\right|\,.
\end{equation}
Here ${\bf v}^{\,\prime}_{ik}={\bf v}_{k}^{\,\prime}-{\bf v}_{i}^{\,\prime}$ and ${\bf v}_{ik}={\bf
v}_{k}-{\bf v}_{i}$ are the relative velocities of two particles of a sort $k$ and $i$ after and before a
collision, and ${\bf e}$ is a  unit vector connecting particles' centers at the collision instant. Note, that
generally $\varepsilon_{ik}$ depends not only on the particular species, involved in an impact, but also on
the relative velocities of the particles, e.g. \cite{book,bshp96,sp98}. In what follows, however, we
will use the simplifying approximation, $\varepsilon_{ik} ={\rm const.}$

The behavior of polydisperse granular gases is rather well studied, not only for a simple binary distribution \cite{wildman,Nut,GarzoDuftyMixture,breyduftyHrenya}, but also for more complex distributions, including
a continuous one.  Namely, polydisperse mixtures with uniform \cite{zippelius} and power-law
\cite{china} size distributions, as well as mixtures of particles of equal mass,
which differ by restitution coefficients~\cite{lambiotte} have been investigated theoretically
\cite{zippelius, lambiotte} and numerically~\cite{zippelius,china,lambiotte}. In
Refs. \cite{GarzohrenyaDuftyI,GarzohrenyaDuftyII} theory of inhomogeneous polydisperse granular gases has
been developed.

Although elaborated theory of granular mixtures allows to compute partial temperatures $T_k$, it may be done
only numerically;  a simple analytical expression which allows to describe distribution of temperatures is
still  lacking. While it is known \cite{GarzoDuftyMixture,zippelius} that for a constant restitution
coefficient the law of equal cooling rates $\dot{T}_k/T_k = \dot{T}_l/T_l$  in a HCS\footnote{For a
velocity-dependent restitution coefficient \cite{book,bshp96} the temperature ratio is not a constant but
demonstrates a complicated non-monotonous evolution, giving rise to super and sub-diffusion in granular
Brownian motion~\cite{annaprl}.} holds true for all $k, \, l$, there is no general relation between
temperature distribution and particle size distribution. It would be rather desirable to have such simple
relation, at least for particular cases.

In the present study we show that if the particle size distribution $n_k(k)$ in a granular mixture of $N$
different species is steep enough, the temperature distribution $T_k \propto m_k^{\alpha} $ is \emph{universal}, that is, the scaling exponent $\alpha$  depends neither on inelasticity of particles nor on
a particular form of the distribution $n_k(m_k)$; it is also (asymptotically) independent  on the total
number of species for $N \gg 1$. Our conclusion holds true for a force-free gas in a HCS as well as for a gas
under a uniform heating, with a power-law dependence of the heating rate on a mass.

\section{Model}
\label{theo} We consider a system with discrete particles' masses: $m_k=m_1k$, where $k=1,\,2, \, \ldots N$,
so that $N$ is the total number of different species. In the context of Planetary Rings, where particles are aggregates built up of smaller grains, $m_1$ may be interpreted as a mass of the smallest particles, hence $k$ characterizes the size of the particles.
%Without the loss of generality we can assume $m_1=1$.
All particles have the same density. For simplicity we assume that the grains are spherical with the diameters of the species $\sigma_k = \sigma_1 k^{1/3}$ and smooth (i.e. the rotational motion is neglected).
We also assume that the number of large particles in the system is significantly smaller than that of small particles, that is, we consider a \emph{steep} particle size distribution. While a quantitative definition of a steep distribution is given below, we wish to emphasize now, that system's kinetics  is mainly determined in this case by small and intermediate particles.  For  space homogeneous granular mixtures addressed here the number densities of species, $n_k(m_k)$,  are  stationary, $\dot{n}_k =0$, and the partial velocity distribution functions $f_k({\bf v}_k,t)$ evolve subjected to the Boltzmann equation:
\begin{equation}
\label{eq:BEgen} \frac{\partial}{\partial t} f_k\left({\bf v}_k, t \right) = I^{\rm res}_{k}+ I^{\rm heat}_k
\, .
\end{equation}
The first term in the right-hand side of Eq.~(\ref{eq:BEgen}) is the collision integral for restitutive
collisions in a multi-component granular gas~\cite{book,GarzohrenyaDuftyI,GarzohrenyaDuftyII}:
\begin{eqnarray}
\nonumber
&&I^{\rm res}_{k}=\sum_{j=1}^{N}\sigma_{kj}^2g_2(\sigma_{kj})\int d{\bf v}_{j} \int d{\bf e} \, \Theta (-{\bf v}_{kj}\cdot {\bf e}\,)\left|{\bf v}_{kj} \cdot {\bf e}\, \right|   \\
&&\times \left[\frac{1}{\varepsilon_{kj}^2} f_k({\bf v}_{k}^{\ \prime\prime},t)f_j ({\bf v}_{j}^{\
\prime\prime},t)-f_k({\bf v}_{k},t)f_j({\bf v}_{j},t)\right] \, , \label{II}
\end{eqnarray}
where $\sigma_{kj}=\left(\sigma_k+\sigma_j\right)/2$ and  $g_2(\sigma_{kj})$ is a contact value of the pair
correlation function for species $j$ and $k$, which takes into account the effects of excluded volume. In
what follows we employ  the  approximation $g_2(\sigma_{kj})=1$, valid for dilute systems. The summation in
Eq.~(\ref{II}) is performed over all species and the rest of the notations has its usual meaning, e.g.
\cite{book}:  ${\bf v}_{k}^{\, \prime\prime}$ and ${\bf v}_{j}^{\, \prime\prime}$ denote pre-collision
velocities in the so-called inverse collision, resulting in the post-collision velocities ${\bf v}_{k}$ and
${\bf v}_{j}$. The Heaviside step-function $\Theta(-{\bf v}_{kj}\cdot{\bf e})$ selects the approaching particles, etc.

The second term $I^{\rm heat}_k$ describes the heating of the system. It quantifies the energy injection into
a granular gas to compensate its losses in dissipative collisions; it is zero for a gas in  a HCS. Here we
consider a uniform heating -- the case when the grains suffer small random uncorrelated  kicks throughout the
volume \cite{WilliamsMacKintosh1996}. To mimic the external driving forces a few types of thermostat have
been proposed \cite{Sant2000}. For a thermostat with a Gaussian white noise, the heating term has the
Fokker-Planck form \cite{vne98}:

\begin{equation}
I^{\rm heat}_k=\frac12\frac{\Gamma_k}{m_k}\frac{\partial^2}{\partial\bf v_k^2}f_k({\bf v_k},t).
\end{equation}
Here the constant $\Gamma_k$ characterizes the strength of the driving force. It may vary for different species, depending on the type of driving \cite{zippelius,Cafiero}. If all species are supplied with the same energy, $\Gamma_k=\Gamma_1=\rm const.$  In the case of the force controlled driving,  $\Gamma_k \propto 1/m_k$,
while in the case of the velocity controlled driving, $\Gamma_k \propto m_k$ \cite{zippelius}. In our study we analyze a more general case of a power-law dependence of $\Gamma_k$ on a particle mass, namely, $\Gamma_k\sim m_k^{\gamma}$ or $\Gamma_k =\Gamma_1 k^{\gamma}$. However, we will not consider the case of driving, based on a local velocity of the granular gas \cite{Cafiero}.

We multiply the Boltzmann Eqs.~(\ref{eq:BEgen}) for $k=1, \ldots N$ with  $m_kv_k^2/2$  and integrate over
$\bf{v}_k$. Using the Maxwellian distribution  with the granular temperatures $T_k(t)$, as a good
approximation~\cite{GarzohrenyaDuftyII,zippelius} for the velocity distribution functions $f_k({\bf
v}_{k},t)$,   we obtain,
\begin{equation}
\frac{d}{dt} T_k= -T_k\sum_{i=1}^{N}\xi_{ki} + \Gamma_k \qquad \qquad k=1,\ldots N \,,
 \label{sys}
\end{equation}
where the cooling rates $\xi_{ik}$ read,
\begin{eqnarray}
\label{eq:xik}
&& \xi_{ki}(t) = \frac{8}{3}\sqrt{2\pi}n_i\sigma_{ik}^{2}g_{2}(\sigma_{ik}) \frac{\mu_{ik}}{m_k} \,
\left(\frac{T_k}{m_k} + \frac{T_i}{m_i} \right)^{1/2}
\\ \nonumber
&&\times \left(1+\varepsilon_{ik}\right)\left[1-\frac{1}{2} \left(1+\varepsilon_{ik}\right)
\frac{\mu_{ik}}{T_k} \left(\frac{T_k}{m_k} + \frac{T_i}{m_i} \right)  \right], 
\end{eqnarray}
with the reduced mass $\mu_{ik} =m_im_k/(m_i+m_k)$. The above expressions for $\xi_{ik}$ coincides with the ones reported in Refs.~\cite{GarzohrenyaDuftyII,zippelius}.

\section{Scaling theory }
\subsection{Homogeneous cooling state}
In a homogeneous cooling state no external energy is supplied to a system. In this case the heating is
lacking, $\Gamma_k=0$, and the evolution equations for the granular temperatures read:
\begin{equation}
\frac{1}{T_k} \frac{dT_k}{dt} = -\sum_{i=1}^{N}\xi_{ki},  \qquad \qquad k=1, \ldots N \,. \label{Ehcs}
\end{equation}
The energy of grains decrease due to dissipative collisions, their motion slows down until all particles come
to a rest. From a general analysis of the Boltzmann equation it may be shown that after a short relaxation time the solution attains a \emph{normal form}, when a distribution function depends on space and time only through hydrodynamic fields. In this regime
the cooling rates $T_k^{-1} dT_k/dt$  for all species with  $k=1,\ldots N$ become equal~\cite{GarzoDuftyMixture,zippelius}, that is,
\begin{equation}
\label{k0} \sum_{i=1}^{N}\xi_{ki} =  G(t) \qquad {\rm for \,~ all}
\qquad {\it k}.
\end{equation}
Here $G(t)$ depends on the relevant characteristics of the system --  concentrations, masses, etc. From Eqs.(\ref{Ehcs})-(\ref{k0}) follows that in the above regime the dependence of $T_k(t)$ on $k$ and $t$  factorizes -- it may be generally written as $T_k(t)= T_1(t)F(k)$ with $F(1)=1$. This also implies the steady-state ratios of granular temperatures $T_k/T_l$ ($k,l=1,\ldots , N$) for a granular gas mixture in a HCS.

Let the number of species in a granular mixture be large, $N \gg 1$, so that the scaling analysis may be applied. Analysing the form of $\xi_{ki}$ in Eq.~(\ref{eq:xik}) we notice that a reasonable assumption for $F(k)$ to fulfil the condition (\ref{k0}) would be   $F(k) \sim m_k^{\alpha}$, or, since $m_k =m_1k$,
\begin{equation}
\label{eq:T_alpha} T_k =T_1 k^{\alpha} \, ,
\end{equation}
and search for the exponent $\alpha$. We focus on the sum $\sum_{i=1}^{N}\xi_{ki}$ and analyze the
consequences of the condition (\ref{k0}).  Approximating  for $N \gg 1$ the summation in Eq.~(\ref{k0}) by
integration, we obtain,
\begin{eqnarray}
\sum_{i=1}^{N}\xi_{ki} \!\!\!\!\!&=& \!\!\! \! a  \!\int_{1}^{N} \! n_i
\!\left(i^{1/3}\!+k^{1/3}\right)^{2}\frac{i}{k+i}
\left(i^{\alpha-1} \! +k^{\alpha-1} \right)^{1/2} \nonumber \\
&\times& \!\!\! \! \left[1-\frac{1}{2} \left(1+\varepsilon\right)\frac{i}{k+i}
\left(1+\left(\frac{i}{k}\right)^{\alpha-1}\right)  \right]  di\label{int0}
\end{eqnarray}
with
\begin{equation}
\label{xi0} a=\frac{2}{3}\sqrt{2\pi}\sigma_1^2
\left(\frac{T_1}{m_1}\right)^{\frac{1}{2}}\left(1+\varepsilon\right)\,,
\end{equation}
where we assume that $\varepsilon_{ik}=\varepsilon={\rm const}$.
Since the condition (\ref{k0})  holds for any $k$, we choose $k \gg 1$. Then, if the particle size-distribution $n_i=n_i(i)$ is steep enough the main contribution to the
integral in Eq.~(\ref{int0}) comes from $i \ll k$. Expanding the integrand in  Eq.~(\ref{int0}) with respect
to $(i/k) \ll 1$ and keeping only the leading terms in the expansion we arrive at
\begin{equation}
\label{eq:xi_gen} \sum_{i=1}^{N}\xi_{ki}   = \left\{
    \begin{array}{ll}
         ak^{\frac{\alpha}{2}- \frac56} \int_1^N i\, n_i \, di & \mbox{if } \, \, \alpha \geq 1 \\
         {} \\
         ak^{- \frac13} \int_1^N i^{\frac{\alpha +1}{2}} \, n_i \, di & \mbox{if } \, \, 0< \alpha < 1 \, .
    \end{array}
\right.
\end{equation}
Here we exclude $\alpha <0$, since it may yield for $i \ll k$ a negative sign for the factor in the square
brackets of the integrand in Eq.~(\ref{int0}). The result of the integration, however, should be positive, as
it gives the cooling rate. For steep distributions $n_i$ one can approximate $N$ in the upper limits of the
integrals in Eq.~(\ref{eq:xi_gen}) by the infinity,
\begin{equation}
\label{eq:steep_cond} \int_1^N i^{p} \, n_i \, di \simeq \int_1^{\infty} i^p\, n_i \, di =\rm const \,,
\end{equation}
where $p=1$ for $\alpha>1$ and $ p=(\alpha +1)/2 $  for $0< \alpha <1$ (see Eq.~(\ref{eq:xi_gen})), so that the sum in (\ref{k0}) does
not (asymptotically, for $N \gg 1$) depend on $N$. Since it neither depends on $k$ [see Eq.~(\ref{k0})], it
follows from Eq.~(\ref{eq:xi_gen}) that $\alpha/2 -5/6=0$. That is,  we conclude that in a HCS \emph{all }
granular gas mixtures with a steep distribution of species size have the same universal power-law
distribution of temperatures (\ref{eq:T_alpha}) with the exponent
\begin{equation}
\label{eq:a} \alpha= \frac53 \approx 1.67 \, .
\end{equation}

\subsection{Heated granular mixtures}
If energy is injected into a granular gas to compensate its losses in dissipative collisions, the system
rapidly settles into a nonequilibrium steady-state \cite{Sant2000,vne98}. In the case of a heated granular
mixtures all granular temperatures $T_k$ attain, after a short relaxation time, constant values, so that
$dT_k /dt =0$. Then Eqs.~(\ref{sys}) read,
\begin{equation}
T_k\sum_{i=1}^{N}\xi_{ki} = \Gamma_1k^{\gamma} \,.\label{Estat}
\end{equation}
We again assume that the size distribution $n_i$ is steep enough, so that in the scaling domain $k \gg 1$ one
can use Eqs.~(\ref{eq:xi_gen})-(\ref{eq:steep_cond}) for the sum $\sum_{i=1}^{N}\xi_{ki}$. Substituting into
the above equation $T_k=T_1k^{\alpha}$, along with Eqs.~(\ref{eq:xi_gen})-(\ref{eq:steep_cond}), we obtain,
taking into account that the exponents of $k$ in the both sides of the equation must be equal:
\begin{equation}
\label{eq:alpha_heat} \alpha    = \left\{
    \begin{array}{ll}
         \frac{5}{9} + \frac{2}{3}\gamma & \mbox{if } \qquad  \gamma \geq \frac23 \\
         {} \\
         \gamma + \frac{1}{3} & \mbox{if } \qquad  -\frac13 \leq \gamma \leq \frac23.
    \end{array}
\right.
\end{equation}
\subsection{Steep size distributions}
The most important for practice are the power-law size distributions $n_k=n_1k^{-\theta}$. These are very common
in nature and industry, as obtained e.g. in fragmentation processes. The condition of a steep distribution
(\ref{eq:steep_cond}) reads for a power-law  distribution,
\begin{eqnarray}\nonumber
&&\theta>2 \qquad \qquad  \quad \rm if  \qquad \alpha>1\\
&&\theta>\frac{\alpha}{2}+\frac{3}{2} \qquad  \, \, \, \, \rm if  \qquad 0<\alpha<1\,. \label{eq:stepcond}
\end{eqnarray}
In the case of the exponential distribution, $n_k=n_1\exp\left(-bk^{\beta}\right)$, the condition of steepness takes in the form: $bN^{\beta}>>1$.

\begin{figure}[ht]
\includegraphics[width=0.9\columnwidth]{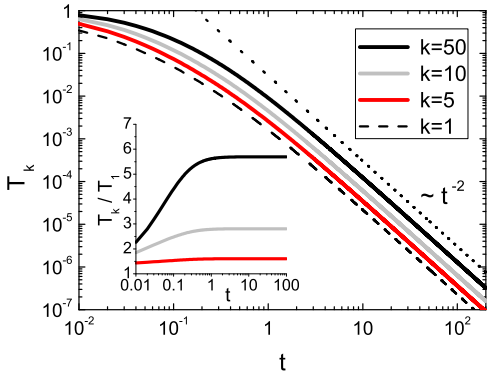}
\caption{The time-dependence of granular temperatures $T_k(t)$ and their ratios $T_k/T_1$ (inset) in a
granular mixture with a power-law size distribution, $n_k  \sim  k^{-3}$, in a HCS state. The
granular temperatures of all species decrease with the same cooling rate according to Haff's law $T_k\sim
1/t^2$ (shown by the dotted line), while the ratios $T_k/T_1$ rapidly evolve to the steady state values (see
the inset). The restitution coefficient $\varepsilon = 0.5$. %Lines from bottom to top correspond to $k=1$, $k=5$, $k=10$, $k=50$.
} \label{GEhcs}
\end{figure}

\begin{figure}[ht]
\includegraphics[width=0.9\columnwidth]{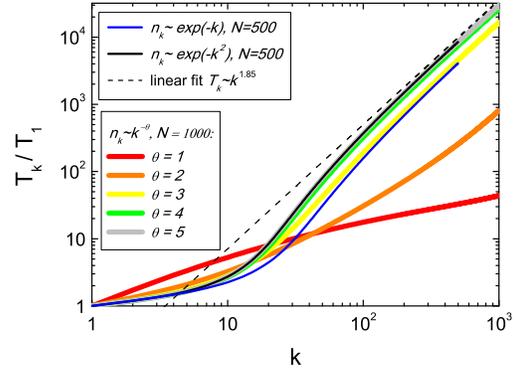}
\caption{The dependence of $T_k/T_1$ on $k=m_k/m_1$ in a granular mixture in a homogeneous cooling state. The
masses of particles are distributed according to the power law $n_k\sim k^{-\theta}$, exponential
distribution $n_k\sim\exp\left(-k\right)$ and Gaussian distribution $n_k\sim\exp\left(-k^2\right)$. It may be
fitted for $k \gg 1$ by the power law $T_k/T_1=k^{\alpha}$, where $\alpha=1.85$ is the same for exponential,
Gaussian and power-law distributions with $\theta>2$. Lines from bottom to top correspond to $n_k \sim
k^{-1}$, $n_k \sim k^{-2}$, $n_k \sim \exp(-k)$, $n_k \sim k^{-3}$, $n_k \sim k^{-4}$, $n_k \sim k^{-5}$,
$n_k \sim \exp(-k^2)$. } \label{Ghcs}
\end{figure}

\begin{figure*}[ht]\includegraphics[width=0.9\columnwidth]{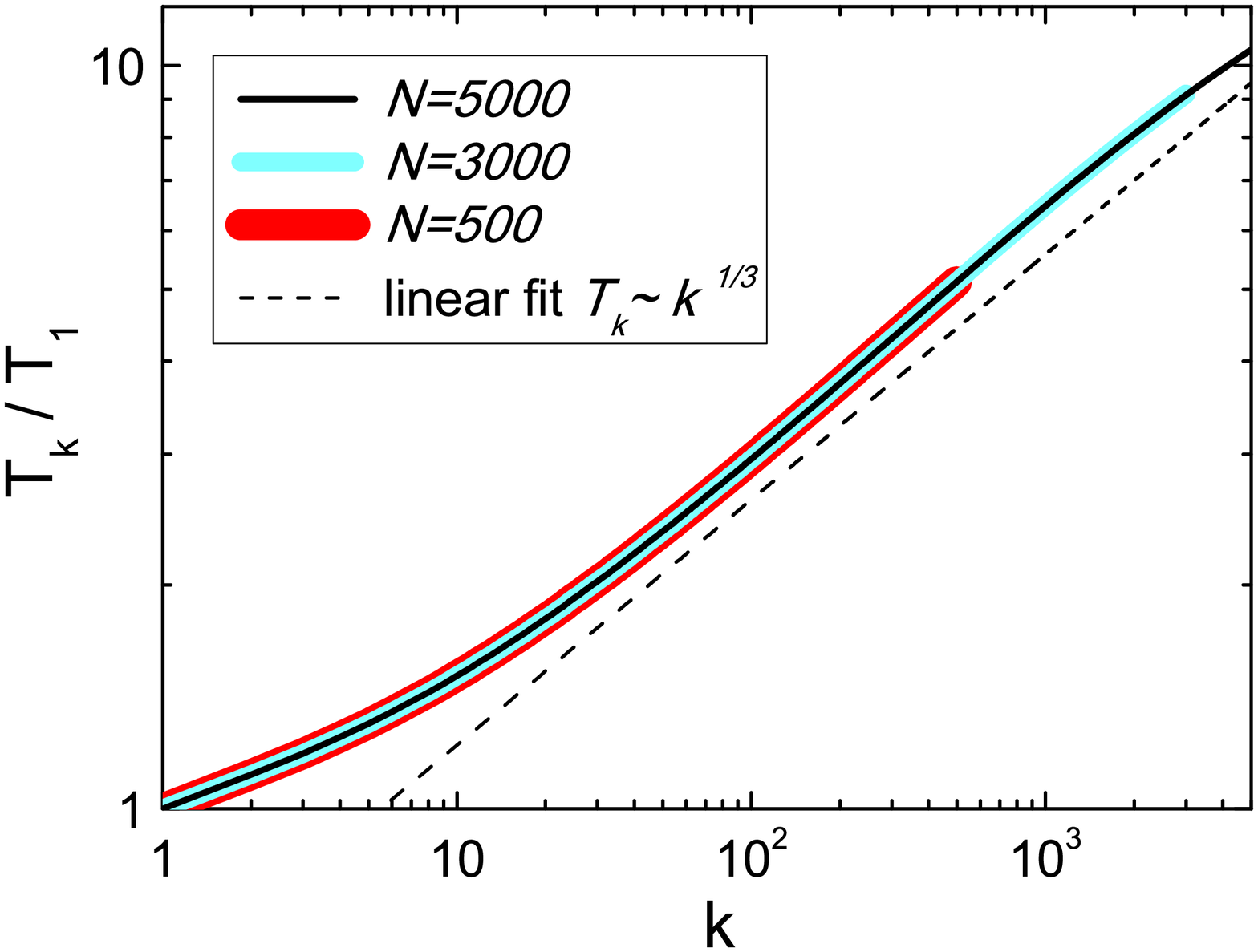}
\includegraphics[width=0.9\columnwidth]{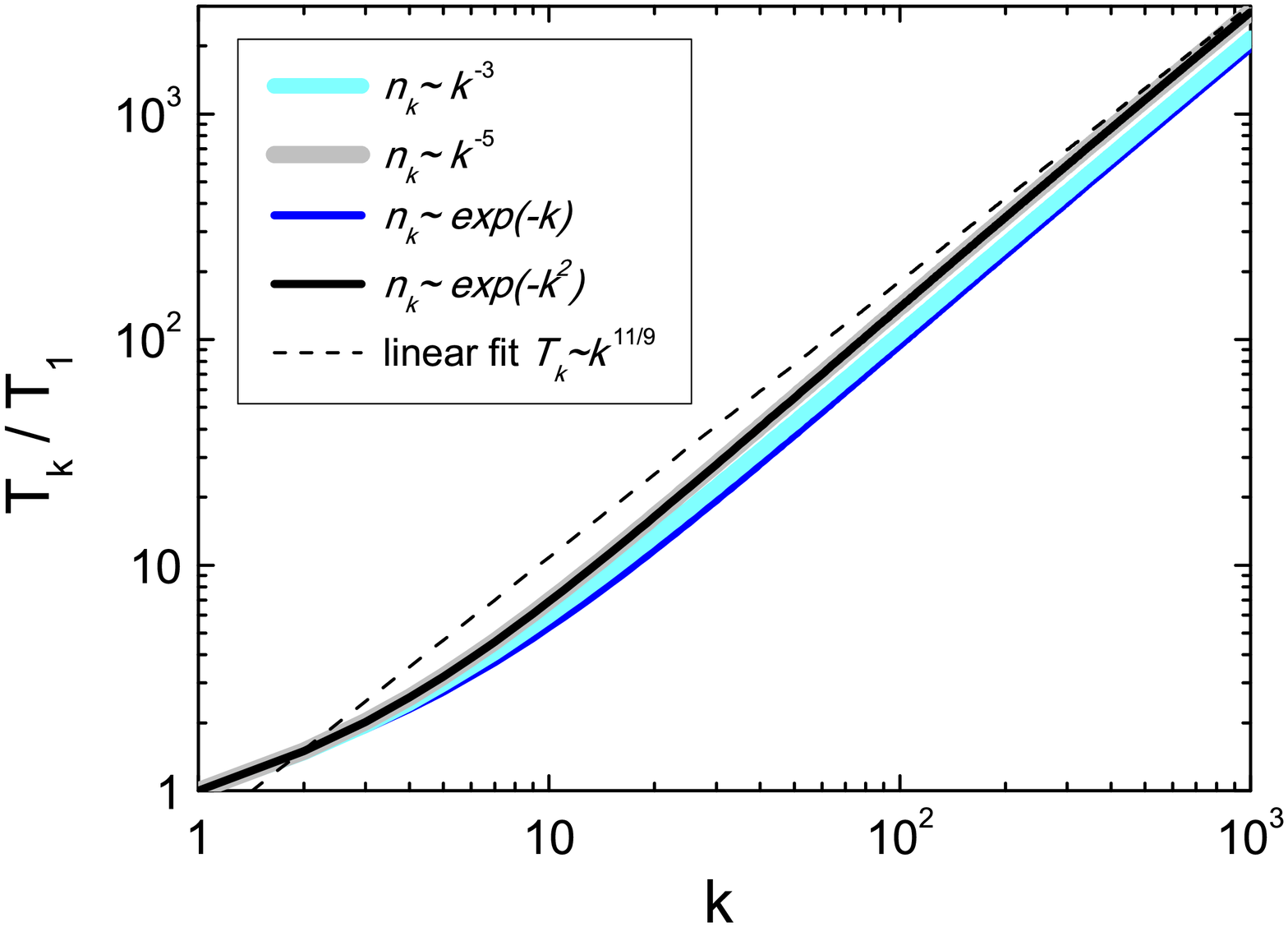}
\caption{The dependence of the ratios of granular temperatures $T_k/T_1$ on the dimensionless particles' mass
$k=m_k/m_1$ for a heated granular mixture. The left panel, $\gamma=0$ (equal energy supply for all
species), $n_k \sim k^{-\theta}$ ($\theta =5 >2$), illustrates independence of the temperature
distribution on the number of species in the system $N$. The right panel, $\gamma=1$ (the velocity
controlled driving), shows independence of the temperature distribution on the particular form of the size
distribution, provided it is steep enough. Lines from the bottom to top correspond to $n_k \sim \exp(-k)$,
$n_k \sim k^{-3}$, $n_k \sim k^{-5}$ and $n_k \sim \exp(-k^2)$. The scaling predictions for the temperature
distribution $ T_k \sim m_k^{\alpha}$ with $\alpha$ from Eq.~(\ref{eq:alpha_heat}) are depicted as the dashed lines, with $\alpha =1/3$ for $\gamma=0$ (left panel)  and $\alpha =11/9$ for $\gamma=1$ (right panel). }
\label{GNmany}
\end{figure*}

\begin{figure*}[ht]
\includegraphics[width=0.9\columnwidth]{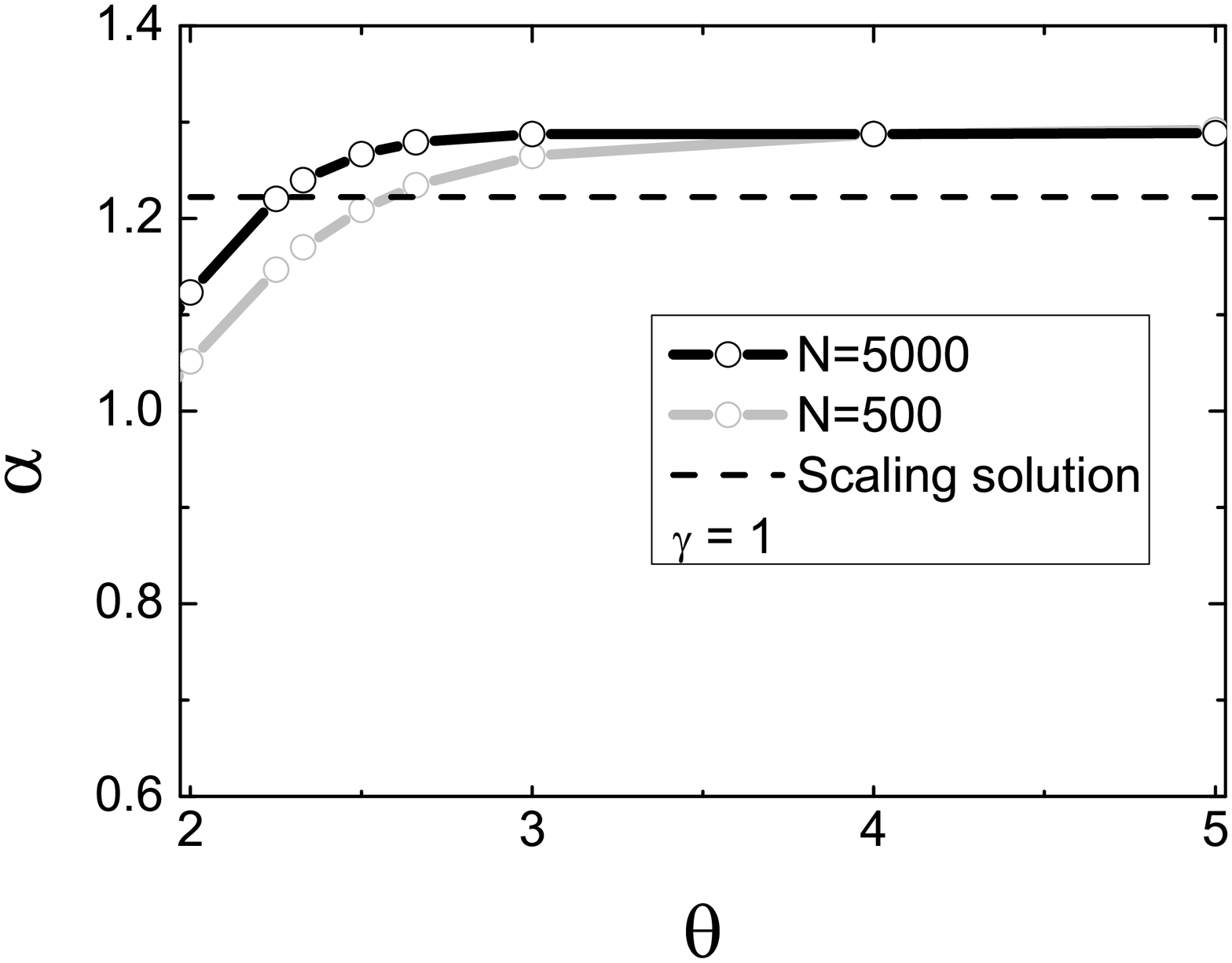}
\includegraphics[width=0.9\columnwidth]{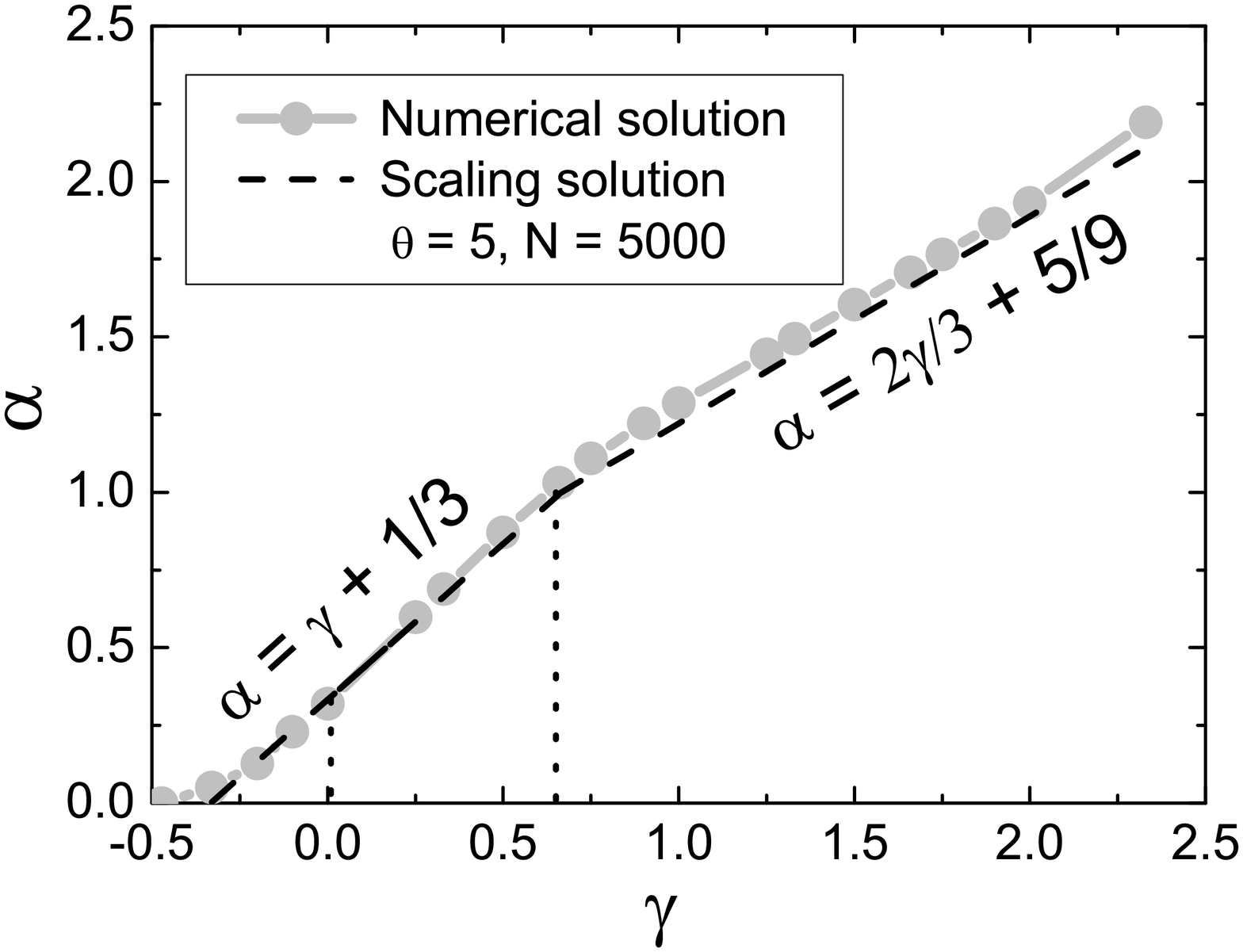}
\caption{The dependence of the exponent $\alpha$ of the granular temperature distribution, $T_k \sim
m_k^{\alpha}$ on the exponent $\theta$ of the particle size distribution, $n_k \sim m_k^{-\theta}$ (left panel)
and on the exponent $\gamma$ of the mass dependence of the heating rate, $\Gamma \sim m_k^{\gamma}$ (right
panel). Thin dashed black  lines show the scaling solution.} \label{Gakg}
\end{figure*}
Note that strictly speaking, the above scaling theory is valid only for $\varepsilon_{ik} = \varepsilon = {\rm const}$. The realistic dependence of $\varepsilon$ on mass and size of particles is however rather weak (see e.g.  \cite{book} (a)) so that the deviations of $(1+\varepsilon_{ik})$ from some average  value are small. Hence, the scaling theory is expected to remain reasonably accurate.

\section{Numerical analysis versus theory}
In order to check the predictions of the scaling theory, we solve numerically the system of differential
equations (\ref{sys}) for granular mixtures in a homogeneous cooling state, as well as in a state under a uniform heating.  In the HCS state the granular temperatures of all species rapidly relax to a state with an equal cooling rate for
all components. All granular temperatures decay under these conditions according to the Haff's law $T_k\sim t^{-2}$
\cite{book}, while the ratios $T_k/T_1$ remain fixed,  Fig.~\ref{GEhcs}. The dependence of $T_k/T_1$ on $k$
in the log-log scale can be fitted for $k \gg 1$  with a straight line, Fig.~\ref{Ghcs}; this confirms our
scaling hypothesis, $T_k \sim m_k^{\alpha}$. The fitting of the numerical solution yields the slope
$\alpha=1.85$. This slope, $\alpha=1.85$, is the same for exponential, Gaussian and power-law distributions
with $\theta>2$. Note however, that the universality of the temperature distribution is violated for $\theta
=2$ and $\theta=1$ -- these size-distributions do not fulfill the condition of steepness (\ref{eq:stepcond}).
The discrepancy between the numerical and scaling result of  $\alpha=1.67$ may be attributed to the systematic error due to replacement of the finite $N$ by the infinite one and the summation by integration.
In the case of a uniform heating, the granular temperatures of all species rapidly evolve to their
steady-state values where  the ratios $T_k/T_1$ are kept constant. As in the case of a HCS the log-log plot
of $T_k/T_1$ on the dimensionless mass $k=m_k/m_1$ demonstrates a linear dependence for the scaling domain,
$k \gg 1$, Fig.~\ref{GNmany}. As it may be seen from Fig.~\ref{GNmany}, the slope of the temperature
distribution does not depend on the number of species in the system $N$ and the particular  form of the size distribution, as the scaling analysis predicts.

We additionally check independence of the temperature distribution on the particle size distribution for the
case of power-law distribution $n_k \sim k^{-\theta}$. In particular, we analyze the threshold value of $\theta$,
above which the scaling theory is valid. In Fig.~\ref{Gakg} (left panel) we show the numerical results for $\gamma=1$. As it may be seen from the figure, the exponent $\alpha$ is indeed
independent on $\theta$, provided it is larger than the threshold $\theta_0>2$, which is close to $2$, as it
follows  from the scaling theory. In the numerical solution we observe that the threshold $\theta_0$ depends
on $N$, tending to the predicted value of $\theta_0 =2$, as $N$ tends to infinity. The numerically found
plateau value of $\alpha = 1.28$ differs from the scaling prediction $\alpha = 11/9=1.22$,
Eq.~(\ref{eq:alpha_heat}), by less than 5\%. Note that for the case of heated gas the replacement in Eq.~(\ref{eq:xik}) of the finite $N$ by the infinite one and of the summation by integration leads to smaller error as compared to the case of the HCS, since in the latter case evolution of granular temperatures are  described by differential equations (\ref{Ehcs}), while in the former one - by an algebraic equations (\ref{Estat})\footnote{As it follows from Eqs.~(\ref{Ehcs}) and (\ref{k0}), $T_k \sim \exp(-\int^t G(\tau) d\tau)$, which implies exponentially strong dependence of the distribution $T_k$ on systematic errors in a HCS.}.

Finally, we study numerically the dependence of $\alpha$ on the power exponent $\gamma$ of the heating rate
$\Gamma_k \sim m_k^{\gamma}$. As it follows from Fig.~\ref{Gakg} (right panel), the scaling theory agrees very
well with the numerical data, which also demonstrate two different slopes on the dependence of $\alpha$ of
$\gamma$: $\alpha =1/3+ \gamma$ for $\gamma <2/3$ and $\alpha =5/9 +2/3 \gamma$ for $\gamma \geq 2/3$.

It is interesting to analyze distribution of granular temperatures $T_k \sim k^{\alpha}$ for some particular
values of  $\gamma$. For the \emph{equal energy} supply for all species, $\gamma=0$, the violation of energy
equipartition nevertheless takes place, with $\alpha = 1/3$. This happens because smaller particles loose in
collisions larger part of their energy than bigger ones. To compensate this "bias" in the collision losses of
small particles, more energy is to be supplied for the smaller grains than to the larger ones, that is, the
exponent $\gamma$ should be negative. Indeed, for $\gamma=-1/3$ the energy equipartition, $T_k={\rm const.}$
is achieved, see Fig.~\ref{Gakg} (right panel).  For another important case of $\gamma=2/3$, the exponent $\alpha$ is unity, $\alpha=1$, and the distribution of the characteristic velocities,  $\bar{v}_k = \left(2 T_k/m_k \right)^{1/2} ={\rm const.}$, is flat. Noteworthy, the distribution of characteristic velocities for different particle sizes in Planetary Rings is seemingly also rather flat \cite{Salo1992}.

%The validity of the scaling theory is based on the condition of the steep size distribution, when
%the number density of small particles significantly exceeds  that of  large particles. Temperature evolution
%in this case is determined mainly by interactions with small particles, while interactions with the larger
%particles may be neglected.
%The steeper the distribution and the larger the number of species, the more accurate the scaling theory is, Fig.~\ref{Gakg} (left panel).

%Unfortunately, we can not compare our theory with the existing MD simulation results, since in spite of vast data for granular mixtures, available in literature, e.g.\cite{zippelius,china,lambiotte}, the according data for a steep size distributions are presently lacking.

\section{Conclusion}
\label{conc}

We have studied by means of a scaling approach and numerically granular gas mixtures with steep size distributions of components, when the number density of small particles significantly exceeds  that of  large particles. We explored  space uniform systems, both in a homogeneous cooling state (HCS) and
under a uniform heating; the latter has been modeled by a white-noise thermostat. We analyzed
mass-dependent heating rates $\Gamma_k$, which depended on species masses as a power-law, $\Gamma_k \sim
m_k^{\gamma}$. We have shown that for \emph{all } steep size distributions, the distribution of granular
temperatures obeys a \emph{universal }  power-law distribution, $T_k \sim m_k^{\alpha}$, where the exponent
$\alpha$ does not depend on inelasticity, number of species and a particular form of the size distribution.
For a HCS we have found numerically the universal exponent  $\alpha =1.85$, which is close to $\alpha =5/3
\simeq 1.67$, predicted by the scaling theory. For heated system we have revealed a piecewise linear
dependence of $\alpha$ on $\gamma$, namely, $\alpha =5/9 +2/3 \gamma$ for $\gamma \geq  2/3$ and $\alpha
=1/3+ \gamma$ for $ 2/3 \geq \gamma \geq  -1/3 $; the predictions of the scaling theory have been confirmed
numerically. The results of our study may be important for  rapid granular flows of highly polydisperse
particles and in studying dynamics and structure of Planetary Rings.

\acknowledgments We thank V. Stadnichuk, M. Tamm, M. Tribelsky, A. Chertovich and Yu. Fomin for fruitful discussions. A.B. gratefully acknowledges the use of the facilities of the Chebyshev supercomputer of the Moscow State University. This work was supported by Russian Foundation for Basic Research (RFBR, project 12-02-31351).

\bibliographystyle{eplbib}
%\bibliography{mix}
%\end{document}

\end{document}